\documentclass[twocolumn,prb,aps,showpacs,floatfix,superscriptaddress]{revtex4}
\usepackage{graphicx}

\newcommand{\cesit}{CeSi$_{2}$}
\newcommand{\cesix}{CeSi$_{x}$}
\newcommand{\cesi}{CeSi$_{1.81}$}

\begin{document}

\title{Suppression of ferromagnetism in CeSi$_{1.81}$ under temperature and pressure}

\author{S. Drotziger}

\affiliation{Physikalisches Institut, Universit\"at Karlsruhe, D-76128 Karlsruhe, Germany}

\author{C. Pfleiderer}

\affiliation{Physikalisches Institut, Universit\"at Karlsruhe, D-76128 Karlsruhe, Germany}

\affiliation{Forschungszentrum Karlsruhe, Institut f\"ur 
Festk\"orperphysik, D-76021 Karlsruhe, Germany}

\affiliation{Physik Department E21, Technische Universi\"at M\"unchen, D-85748 Garching, Germany}

\author{M. Uhlarz}

\affiliation{Physikalisches Institut, Universit\"at Karlsruhe, D-76128 Karlsruhe, Germany}
  
\author{H. v. L\"ohneysen}

\affiliation{Physikalisches Institut, Universit\"at Karlsruhe, D-76128 Karlsruhe, Germany}

\affiliation{Forschungszentrum Karlsruhe, Institut f\"ur 
Festk\"orperphysik, D-76021 Karlsruhe, Germany}
  
\author{D. Souptel}

\affiliation{Leibniz-Institut f\"ur Festk\"orper- und Werkstoffforschung Dresden, PF 27\,01\,16, D-01171 Dresden, Germany}

\author{W. L\"oser}

\affiliation{Leibniz-Institut f\"ur Festk\"orper- und Werkstoffforschung Dresden, PF 27\,01\,16, D-01171 Dresden, Germany}

\author{G. Behr}

\affiliation{Leibniz-Institut f\"ur Festk\"orper- und Werkstoffforschung Dresden, PF 27\,01\,16, D-01171 Dresden, Germany}

\date{\today}

\begin{abstract}
We have studied the pressure dependence of the magnetization of single crystalline CeSi$_{1.81}$. 
At ambient pressure ferromagnetism develops below $T_{\rm{C}} = 9.5$\,K.
Below $\sim5$\,K an additional shoulder in low-field hysteresis loops and a metamagnetic crossover around 4\,T suggest the appearance of an additional magnetic modulation to the ferromagnetic state.
The suppression of the magnetic order in {\cesi} as function of temperature at ambient pressure and as function of pressure at low temperature are in remarkable qualitative agreement.
The continuous suppression of the ordered moment at $p_c\approx13.1$\,kbar suggests the existence of a ferromagnetic quantum critical point in this material.
\end{abstract}

\maketitle

\vskip2pc

%

\section{Introduction}
Quantum phase transitions have recently attracted great scientific interest.
In contrast to phase transitions at finite temperatures where thermal order-parameter fluctuations are dominant, quantum phase transitions, e.g., phase transitions at $T = 0$ controlled by a non-thermal parameter like pressure or magnetic field, are driven by quantum fluctuations.
Magnetic quantum phase transitions (QPT) involving the formation of a spin density wave out of an itinerant electron system have been studied theoretically for quite a long time. \cite{0a,0b,0c}
However, experiments on intermetallic Ce and Yb compounds undergoing an antiferromagnetic quantum phase transition defy a simple explanation, notably the systems CeCu$_{6-x}$Au$_x$ \cite{0d,0e} and YbRh$_2$(Si$_{1-x}$Ge$_x$)$_2$ \cite{0f,0g} which have been studied extensively.
The ferromagnetic to paramagnetic quantum phase transition of itinerant electron systems is considered to be the simplest quantum phase transition of the conduction electrons in metals.
To date all pure materials in which ferromagnetic quantum phase transitions have been studied, e.g. UGe$_2$ \cite{pfle02}, ZrZn$_2$ \cite{uhl04} and Ni$_3$Al \cite{nik05} display a behavior that can be well explained in terms of first-order behavior.
On the theoretical side, several mechanisms have been identified that pre-empt ferromagnetic quantum criticality by a first-order transition. \cite{voj01}


The search for ferromagnetic quantum phase transitions focused so far on transition-metal compounds like MnSi \cite{pfle97}, ZrZn$_2$ \cite{uhl04} and Ni$_3$Al 
\cite{nik05}.
Among $f$-electron materials, in particular the ferromagnetic superconductors UGe$_2$ \cite{pfle02}, URhGe \cite{har05} and UIr \cite{aka04} have attracted interest.
In Ce-based ferromagnets, the zero temperature phase transition from a ferromagnetic to an antiferromagnetic state has been studied in CeRu$_2$Ge$_2$. \cite{sue99}
For CePt \cite{lar05} the ferromagnetic quantum phase transition has been studied.
However, no measurements of the magnetization being the order parameter were presented, leaving many questions unanswered.

In {\cesix}, ferromagnetism, antiferromagnetism and paramagnetism have been reported to exist within a wide homogeneity range.
Brauer and Haag \cite{bra01} reported that stoichiometric CeSi$_2$ crystallizes in the $\alpha$-ThSi$_2$ structure.
The wide homogeneity range  $1.78 < x < 2$ that makes {\cesix} of interest in our study was first noticed by Ruggiero \textit{et al.} \cite{rug01}
Yashima \textit{et al}. \cite{yas01} inferred from the specific heat $C(T)$ and the magnetic susceptibility $\chi(T)$ of CeSi$_2$ and CeGe$_2$ that {\cesit} is a nonmagnetic intermediate-valent system with a metallic Fermi-liquid groundstate down to 100\,mK. 
In contrast, CeGe$_2$ displays a magnetic transition at $T_{\rm{C}}=7$\,K, where the susceptibility suggested ferromagnetic order.
Electron diffraction gave evidence for a superlattice structure in CeGe$_2$, while no such feature was observed in {\cesit}.

Further studies \cite{yas02,yas03} of CeSi$_x$ for $x = 1.7, 1.8, 1.85, 1.9, 2$ suggested an even wider homogeneity range.
At the same time, a ferromagnetic groundstate was observed below $T_{\rm{C}}\approx10$\,K for $x = 1.8$ and a nonmagnetic intermediate-valence state for $1.85 < x < 2$. 
The spontaneous ferromagnetic moment of $\mu_{\rm{S}}\approx0.3$ $\mu_{\rm{B}}$(Ce-atom)$^{-1}$ is strongly reduced by comparison to the free Ce$^{3+}$ moment of $\mu_{\rm{free}} = 2.54$ $\mu_{\rm{B}}$(Ce-atom)$^{-1}$, which is attributed to Kondo screening in a lattice.
Hence, {\cesix} has been referred to as a ferromagnetic dense Kondo system.

The magnetic-nonmagnetic boundary as function of Si content was investigated in various studies.
Measurements of $C(T)$ and $\chi(T)$ \cite{yas04} indicated a divergent Sommerfeld constant $\gamma = C_{\rm{V}}/T$ in the Si concentration range $1.8 < x < 1.85$.
Sato \textit{et al.} \cite{sat01} reported measurements of the electrical resistivity and the paramagnetic susceptibility for the $a$- and $c$-axis of a CeSi$_{1.86}$ single crystal, i.e. near the ferromagnetic instability. 
The lattice parameters of the crystal grown by a floating zone technique were $a = 4.182$ \AA\,\, and $c = 13.85$ \AA.
Strongly anisotropic behavior is observed in $\rho(T)$ and $\chi(T)$. 
At low temperatures both quantities show a $T^2$ dependence that indicates Fermi-liquid behavior. 
Based on the thermal expansion and magnetic susceptibility of two CeSi$_{x}$ single crystals ($x = 1.70$ and $x = 1.86$) Sato \textit{et al.} \cite{sat02} suggested a  $\Gamma_7$ crystal-field groundstate.
Measurements of the magnetization established that the $c$-axis is the magnetically hard axis. 
For the easy axis the spontaneous magnetization at $T = 4.2$\,K is given by $\mu_{\rm{S}} \approx 0.44$ $\mu_{\rm{B}}$(Ce-atom)$^{-1}$.
At magnetic fields $H_{\rm{c1}} = 30$\,kOe and $H_{\rm{c2}} = 47$\,kOe anomalous behavior  was observed, where measurements of the magneto-caloric effect for magnetic field along the $a$-axis was interpreted as providing a weak antiferromagnetic modulation.
Later Sato \textit{et al.} \cite{sat03} also reported measurements of the specific heat $C_{\rm{V}}(T)$, electrical resistivity $\rho(T)$ and magnetic susceptibility $\chi(T)$ of two {\cesix} single crystals grown by an inductive floating zone method with the same Si concentrations.
To determine the Si content, the specific heat was compared with results for polycrystalline samples. 
The analysis of $\chi(T)$ conjectured the ground state to be a $\Gamma_7$-like doublet, as also suspected earlier. \cite{sat02}
$\rho(T)$ was found to exhibit a broad maximum indicating the onset of coherence. 

Shaheen \textit{et al.} \cite{sha01} reported a transition from the tetragonal $\alpha$-ThSi$_2$ structure to an orthorhombic $\alpha$-GdSi$_2$ phase for low Si concentrations $x\leq1.84$.
For $x < 1.84$ ferromagnetic order was observed with evidence for further magnetic components. 
At ambient and high pressure the spontaneous magnetic moment was found to be strongly reduced as compared to the free Ce$^{3+}$ moment.
This reduction was attributed to strong Kondo screening.
Precision measurements of $\rho(T)$ on polycrystalline CeSi$_{x}$ samples in the range $1.6 < x < 1.9$ were also made by Lee \textit{et al.} \cite{lee01}

In neutron scattering experiments on a single crystal with $x = 1.8$ ($T_{\rm{C}} = 13.4$\,K), a spin-wave-like response was observed for $T < T_{\rm{C}}$ over the entire Brillouin zone with abnormally large line widths both above and below $T_{\rm{C}}$.\cite{koh01}
This suggests that the magnetic excitations are unstable at low $T$ because of the Kondo effect.

Further measurements of $\rho(T)$ and $\chi(T)$ on single crystals indicated strong anisotropies and confirmed the $(ab)$-plane as easy magnetic plane.\cite{pie01} 
For $x = 1.71$ magnetic order was observed below $T_{\rm{C}}= 12.5$\,K, while no magnetic order could be determined in a sample with $x = 1.86$.
In the ordered system a trend toward Kondo behavior is suggested because the magnetic moment $\mu_{\rm{S}}\approx0.47$ $\mu_{\rm{B}}$(Ce-atom)$^{-1}$ is strongly reduced compared to the free Ce moment  $\mu_{\rm{free}} = 2.54$ $\mu_{\rm{B}}$(Ce-atom)$^{-1}$.

The influence of annealing CeSi$_{x}$ in the range of $1.7 < x < 1.84$ was reported by Shaheen \textit{et al.} \cite{sha03}
The ac susceptibility $\chi_{\rm{ac}}(T)$ in the range 10\,K to 15\,K shows, depending on annealing conditions, either a single peak or multiple peaks, which are attributed to multiple magnetic transitions.
This suggests that the magnetism of this system is not that of a simple ferromagnet, but has been ascribed to two different types of magnetic (ferromagnetic and antiferromagnetic) order.
However it is also suggested that a non-uniform distribution of Si-vacancies can lead to this complex behavior.

Although {\cesix} has been considered as a system with broad homogeneity range recently Souptel \textit{et al.} \cite{sou01} successfully grew comparatively large single crystals of high perfection within the narrow region $1.81<x<1.82$ only, using float zoning with an image furnace.
For the optically float-zoned samples the structural phase transition is found at $x = 1.85$.
This is exactly the Si concentration of the paramagnetic to ferromagnetic phase transition, thus clarifying the relation between structure and magnetic order.

In the light of these new results, we have revisited the question of the ferromagnetic to paramagnetic transition in {\cesix} in a detailed study of the magnetic properties of {\cesi} as function of pressure in order to compare the suppression of ferromagnetism as function of temperature with the suppression as function of pressure.
The magnetic anisotropy is associated in an obvious manner to the structural anisotropy.
Both anisotropies are rather strong, in particular the magnetic anisotropy field is very large.
We measured the magnetization only along the $a$-axis of the single crystal and hence cannot draw a conclusion about the anisotropy within the easy plane.
But we feel safe to assume that the applied pressures do not tilt the easy axis away from the $(ab)$-plane since it is unlikely that the underlying orbital orientation of the Ce atoms, which drives the magnetic anisotropy, will change for the very weak changes of lattice constants imposed by the pressures applied in our study.
A direction within the easy magnetic plane was therefore chosen to obtain information of the pressure dependence of magnetic properties.

\section{Experimental technique}
Single crystals of {\cesi} were grown at IFW Dresden with an image furnace as described previously. \cite{sou01} 
Cylindrical feed-rods were prepared from high-purity Ce ($>99.9\,\%$) and Si ($>99.99\,\%$) on a water-cooled copper hearth.
The Si content of the resulting single crystals was determined from the unit-cell volume as measured by powder $x$-ray diffraction and compared with standards of known concentration.
The crystal structure for $x=1.82$ was carefully studied by four-circle $x$-ray diffraction, and an incommensurate superstructure was suggested to occur. \cite{lei05}
Neutron scattering experiments on the same samples do not confirm these findings, suggesting that the superstructure is a surface property. \cite{jano01}
Samples were cut with a diamond wire saw from the ingot.
For our high-pressure magnetization study we used a rectangular sample cut along the $a$-axis. 
The lattice parameters for $x = 1.81$ are $a = 4.173$ \AA\,\, $b = 4.181$ \AA\,\, and $c = 13.846$ \AA.

The magnetization was measured with a vibrating sample magnetometer (VSM) at temperatures down to 1.5\,K under magnetic field $B$ up to 12\,T at the University of Karlsruhe.
The magnetization as function of pressure $p$ up to 17\,kbar was measured with a bespoke non-magnetic Cu:Be clamp cell.
The empty pressure cell was measured separately and the signal corrected accordingly.
The pressure transmitting medium was a methanol:ethanol mixture of 4:1 volume ratio.
The pressure was determined from the superconducting transition of Sn, after carefully demagnetizing the superconducting solenoid of the VSM.

\section{Results}
The presentation of the results of our high-pressure magnetization study proceeds as follows.
We first focus on the effect of temperature at ambient pressure. 
Here the magnetic order 'melts' with increasing temperature.
Next we present the suppression of magnetic order under high pressure.
This addresses the 'melting' of magnetic order as function of a non-thermal control parameter.
Finally, we present the temperature dependence at high pressure, i.e., changes of thermally 'melting' magnetic order at various pressures.

\subsection{Ambient pressure}

\begin{figure}[h]
\includegraphics[width=0.4\textwidth]{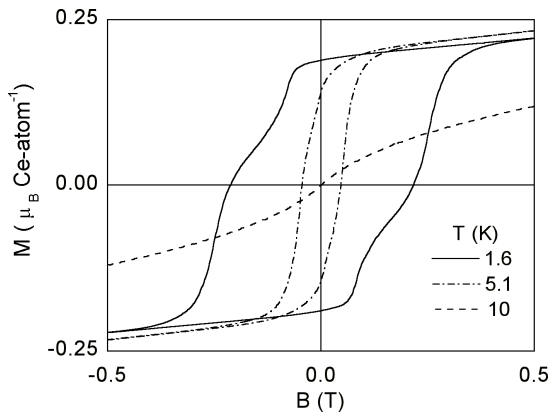}
\caption{Field dependence of the magnetization at typical temperatures $T = 1.6, 5.1, 10$\,K. The magnetic field $B$ as applied $\parallel\, \langle 100 \rangle$ was varied between -\,0.5\,T\,$\rightarrow$\,0.5\,T\,$\rightarrow$\,-\,0.5\,T. The widest hysteresis loop for $T = 1.6$\,K shows a small shoulder which vanishes at 5 K.}
\label{figure1}
\end{figure}

\begin{figure}[ht]
	\includegraphics[width=0.39\textwidth]{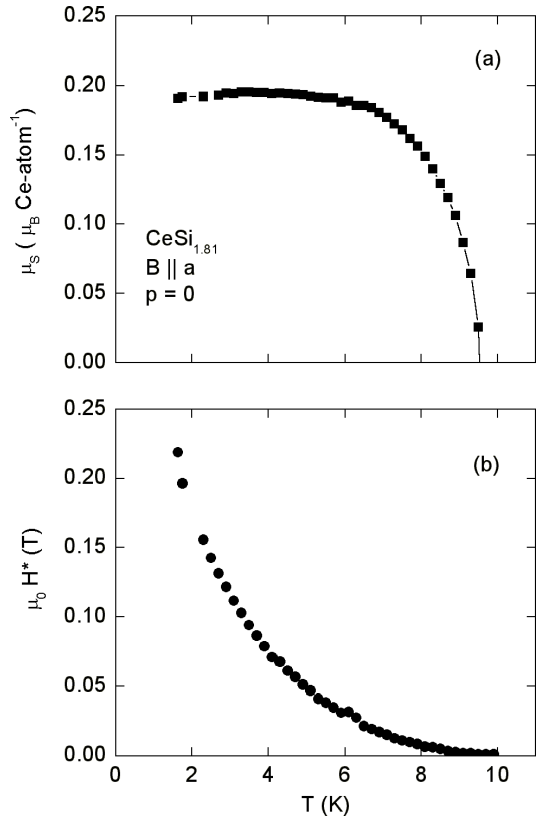}
	\caption{(a) Ordered magnetic moment $\mu_{\rm{S}}$ as function of temperature $T$ for $B\parallel\,\langle 100 \rangle$. The data points were obtained by extrapolation of the hysteresis loops to $B=0$. The line connecting the points serves as guide to the eye. (b) Temperature dependence of the coercive field $H^*(T)$ at ambient pressure as extracted from magnetic hysteresis loops between -\,0.5\,T\,$\rightarrow$\,0.5\,T\,$\rightarrow$\,-\,0.5\,T.}
	\label{figure2}
	\end{figure}

Shown in Fig.~\ref{figure1} are typical hysteresis loops in the range -\,0.5\,T\,$\leq B\leq$\,0.5\,T for three characteristic temperatures $T = 1.6$\,K, $5.1$\,K and $10$\,K, i.e., well below $T_{\rm{C}}$, near $T_{\rm{C}}$ and above $T_{\rm{C}}$, respectively.
The shape of the magnetic field dependence down to $T^*\approx5$\,K ($\approx T_{\rm{C}}/2$) is characteristic of a purely ferromagnetic state with a small ordered moment.
Above a certain coercive field the sample is in a single domain state.
Below $T^*$ a shoulder emerges as an additional feature that suggests an antiferromagnetic component on top of the otherwise ferromagnetic signal.
We have extracted the ordered magnetic moment $\mu_{\rm{S}}$ from magnetization loops by extrapolating $B\to0$ from field values that are clearly above the coercive field.
%
%
While due to nonlinearity in the Arrott plots, these cannot be employed unambiguously to extract $T_{\rm{C}}$, such an extraction, where possible, leads to the same results as that of the saturation moment.
The ordered moment as function of temperature is shown in Fig.~\ref{figure2}a.
For $T = 1.7$\,K, the lowest temperature studied, $\mu_{\rm{S}}\approx\,0.18$ $\mu_{\rm{B}}$(Ce-atom$^{-1}$).
With increasing temperature $\mu_{\rm{S}}(T)$ passes over a shallow maximum at 4\,K which is close to $T^* = 5$\,K.
$\mu_{\rm{S}}(T)$ vanishes continuously at $T_{\rm{C}} = 9.5$\,K characteristic of a second-order phase transition.

The shallow maximum of  $\mu_{\rm{S}}(T)$ and the emergence of the additional shoulder in the magnetization loops signal the presence of an additional magnetic component on top of the otherwise ferromagnetic signal below $T^*$.
However, the coercive field $H^*$ as function of temperature does not exhibit any pronounced features as shown in Fig.~\ref{figure2}b.
While the weak structure around 6\,K is within the error bar, an influence of the onset of the change of the hysteresis loops (Fig.~\ref{figure1}) cannot be excluded.

\begin{figure}
	\includegraphics[width=0.4\textwidth]{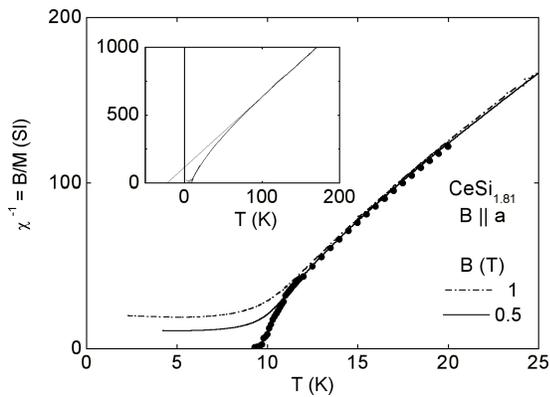}
	\caption{Temperature dependence of the inverse dc susceptibility $\chi^{-1}(T)$ for low fields ($B = 0.5, 1$\,T) applied along the $a$-axis of the CeSi$_{1.81}$ single crystal. In addition, the inverse differential susceptibility $\chi^{-1}_{\rm{d}}(T)$ (circles) extrapolated from hysteresis loops above $T_{\rm{C}}$ is displayed. Near $T_{\rm{C}}$ an additional downward curvature appears in both data sets. The inset shows the inverse dc susceptibility $\chi^{-1}(T)$ at high temperatures together with an extrapolation to $\chi^{-1}(T)=0$.}
	\label{figure3}
	\end{figure}

The Curie temperature $T_{\rm{C}}$ is normally defined directly from the order parameter, i.e., $\mu_{\rm{S}}\to0$.
Alternatively, it may be defined from $\chi^{-1}(T_{\rm{C}})\equiv0$. 
The dc susceptibility $\chi\approx M/B$, has been determined at $B = 0.5$\,T and $B = 1$\,T, while the differential susceptibility $\chi_{\rm{d}}(T)= {\rm{d}}M/{\rm{d}}B$ has been extracted from low-field cycles above $T_{\rm{C}}$ for $B > 0$, corresponding to a measurement for $B\rightarrow0$ (see Fig.~\ref{figure1} for an example). 
Shown in Fig.~\ref{figure3} is a comparison of the inverse dc susceptibility $\chi^{-1}(T)$ and the differential inverse susceptibility $\chi^{-1}_{\rm{d}}(T)$ (full dots) for temperatures below $T = 25$\,K. 
$\chi^{-1}_{\rm{d}}(T)$ drops continuously to zero for $T_{\rm{C}} = 9.5$\,K in excellent agreement with the temperature below which $\mu_{\rm{S}}$ emerges, whereas the inverse dc susceptibility reaches a finite constant value corresponding to a finite polarization in the external magnetic field as can be directly seen in Fig.~\ref{figure1}.

Far above $T_{\rm{C}}$ the temperature dependence of $\chi$ is Curie-Weiss-like with an effective moment $\mu_{\rm{eff}} = 2.05$ $\mu_{\rm{B}}$(Ce-atom$^{-1}$) as shown in the inset of Fig.~\ref{figure3}.
The slight reduction of the effective moment as compared with the free-ion value, $\mu_{\rm{free}} = 2.54$ $\mu_{\rm{B}}$(Ce-atom$^{-1}$), may be related to the presence of crystal electric fields and the magnetic anisotropy.
Alternatively, a phenomenological interpretation in terms of a spin fluctuation model \cite{mor01} may also be possible.
However, the negative curvature of $\chi^{-1}(T)$ below $\sim100$\,K speaks against this possibility.
In comparison to the effective moment, the ordered moment $\mu_{\rm{S}}$ is reduced by over an order of magnitude. 
In intermetallic Ce compounds this is normally taken as a signature characteristic of the Kondo effect.

Above $\sim100$\,K the extrapolated Curie-Weiss temperature is given by $\Theta \approx -20$\,K. 
This suggests effective antiferromagnetic interactions between the magnetic moments at high temperatures and may signal that the system is in fact a ferrimagnet (or "canted ferromagnet") as already inferred from the magnetization and saturation moment. 
For $T < 100$\,K the inverse susceptibility curves downwards and differs from the Curie-Weiss dependence observed at high temperatures.
The downward curvature below $\sim$\,20\,K may signal the effect of low-lying crystal electric fields.
It may be a fortuitous coincidence that the inverse susceptibility above 20\,K, when extrapolated to zero, vanishes around 5\,K, i.e., the temperature $T^*$ below which the magnetization loops display the additional shoulder.

\begin{figure}[ht]
	\includegraphics[width=0.4\textwidth]{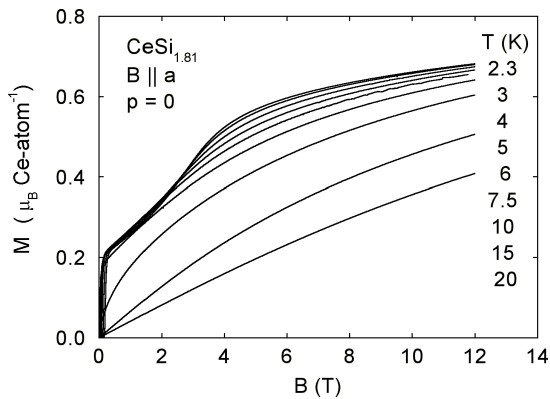}
	\caption{Magnetic field dependence of the magnetization $M(B)$ up to 12 T at different temperatures. The magnetic field was directed along the $a$-axis ($ B\parallel\,\langle 100 \rangle $) of the CeSi$_{1.81}$ single crystal. The magnetization decreases monotonically with increasing temperatures.}
	\label{figure4}
	\end{figure}

We next turn to the magnetic properties at ambient pressure under high magnetic fields.
Shown in Fig.~\ref{figure4} is the easy-plane magnetization of single-crystal CeSi$_{1.81}$ as function of magnetic field up to 12\,T for temperatures in the range $2.3 < T < 20$\,K. 
%
Below $T\lesssim10$\,K the magnetization displays an initial jump to a finite value for $B \rightarrow 0$ consistent with ferromagnetic order.
With increasing field up to $12$\,T the magnetization continues to grow without apparent bound.
At the lowest temperatures studied it reaches $\mu\approx\,0.65$ $\mu_{\rm{B}}$(Ce-atom$^{-1}$), which is a small fraction of the effective Curie-Weiss moment and the free-ion value.
While the reduced high-field magnetization may be the consequence of low-lying crystal electric fields and/or the Kondo effect, it is also taken as the typical signature of weak itinerant magnetism in transition metal compounds.
As an unusual feature the magnetization displays an additional increase around $B\approx4$\,T.
An inflection point of the magnetization as function of magnetic field is now commonly referred to as indication of metamagnetism.
Such a metamagnetic transition has been observed in several heavy-fermion systems without magnetic oder, e.g. CeRu$_2$Si$_2$ \cite{pau01} or CeCu$_6$.\cite{Oh}
A possible interpretation for our data, as for the case of nonmagnetic CeRu$_2$Si$_2$ or CeCu$_6$, would be the weakening of the Kondo effect in high magnetic fields.
The new feature here is that this transition exists "above" a ferromagnetic groundstate.
The inflection point vanishes for temperatures above $T^*\approx5$\,K.
It will be a challenge to establish a possible link of the shoulder in the low-field hysteresis loops and the inflection point in the high-field magnetization.
Indeed, the features in the hysteresis loop are strongly tied to the hump in $M(B)$ since both vanish at the same temperature or pressure (see below). 
A definite possibility is therefore a ferrimagnet arising from different Ce sublattices where a large temperature or pressure renders the two sites equivalent. 
Two different Ce lattice sites might arise from the nonstoichiometric Si concentration leading to differences in the atomic Ce environment.

\subsection{High pressure}

\begin{figure}[hb]
	\includegraphics[width=0.4\textwidth]{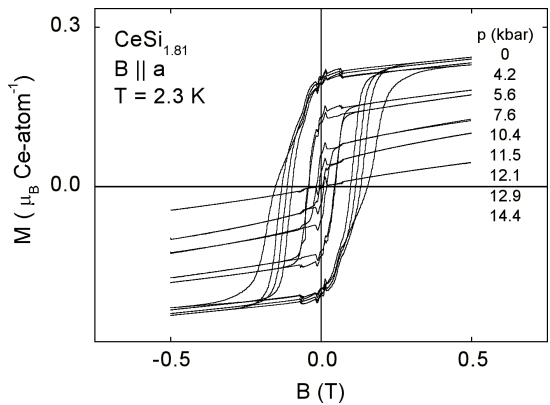}
	\caption{Hysteresis loops between -\,0.5\,T$\rightarrow$\,0.5\,T$\rightarrow$\,-\,0.5\,T for different pressures at $T = 2.3$\,K. The field was orientated along the $a$-axis of the single crystal as for $p = 0$. The small spikes and humps are remains of the Sn gauge. Applied pressures lie in the range $0 < p < 14.4$\,kbar. The high-field magnetization decreases monotonically as the pressure is increased from 7.6 to 14.4\,kbar. For 4.2\,kbar and 5.6\,kbar $M$ increases slightly (see also Fig.~\ref{figure6}b).}
	\label{figure5}
        \end{figure}

	\begin{figure}[hb]
	\includegraphics[width=0.4\textwidth]{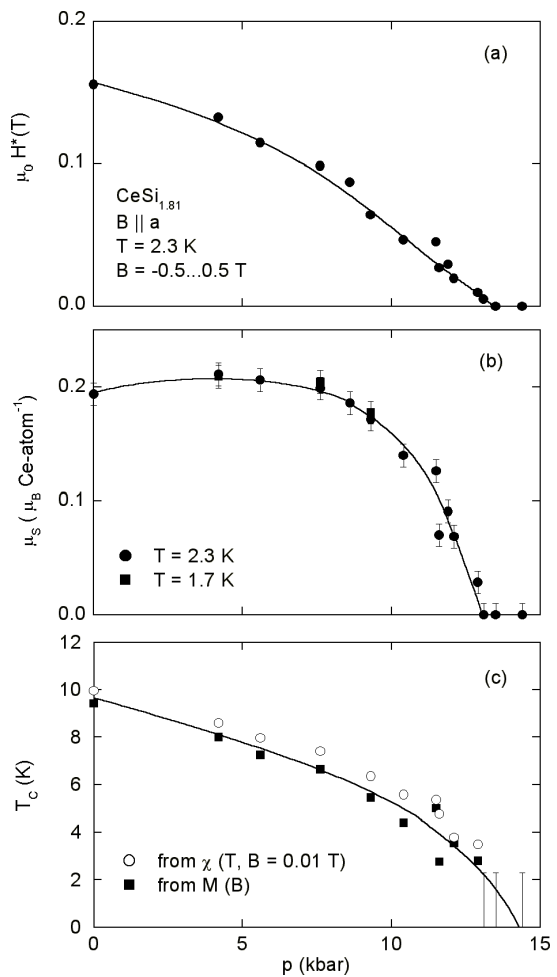}
	\caption{(a) Pressure dependence of the coercive field $H^*(p)$ extracted from hysteresis loops between -\,0.5\,T$\rightarrow$\,0.5\,T$\rightarrow$\,-\,0.5\,T at $B(M=0)$ for different pressures at $T = 2.3$\,K. (b) Spontaneous magnetic moment $\mu_{\rm{S}}$ as function of pressure at $T = 2.3$\,K and $T = 1.7$\,K. (c) Curie temperature $T_{\rm{C}}$  as function of pressure $T = 2.3$\,K. As for $p = 0$ the field was directed along the $a$-axis of the single crystal. Applied pressures lie in the range $4.2$\,kbar  $< p < 14.4$\,kbar.}
	\label{figure6}
	\end{figure}
We now present the pressure dependence of the magnetization for magnetic field applied along the $a$-axis, i.e. the (zero-pressure) easy magnetic axis.
As pressure suppresses ferromagnetism we proceed in correspondence with the presentation of the data as function of temperature.
Data were recorded for the same combination of temperature and field sweeps as studied at ambient pressure.

We begin by discussing typical magnetization loops in the range -\,0.5\,T$\rightarrow$\,\-0.5\,T$\rightarrow$\,-\,0.5\,T as measured for temperatures in the range $2.3$\,K\,$ < T < 15$\,K at various pressures. 
Data at $T = 2.3$\,K for different pressures are shown in Fig.~\ref{figure5}.
At ambient pressure the hysteresis loop is characterized by the coercive field and the shoulder as already shown above.
With increasing pressure the ordered magnetic moment and the size of the coercive field decrease monotonically and vanish for pressures above $\sim13$\,kbar (Fig.~\ref{figure6}a).
Small sharp features in the $M(B)$ data and a tiny residual offset for $B\to0$ at the highest pressures are attributed to spurious effects and show the level to which the signal of the empty pressure cell, notably the contribution of the superconducting piece of Sn, could be subtracted.

The evolution of the ordered magnetic moment at $T=2.3$\,K as function of pressure, $\mu_{\rm{S}}(p)$, is shown in Fig.~\ref{figure6}b.
With increasing pressure, $\mu_{\rm{S}}(p)$ increases slightly and displays a shallow maximum for $p^*\approx5$\,kbar before dropping off gradually and vanishing continuously above an extrapolated critical pressure $p_c=13.1$\,kbar.
There is a remarkable qualitative analogy of the suppression of the ordered moment under temperature and pressure, respectively, that is not expected.
The behavior suggests that temperature and pressure have the same qualitative effect on the magnetic properties.
In particular, the similarity suggests that the critical point at $p=0$ as function of temperature $T \rightarrow T_{\rm{C}}$ has a counterpart in a quantum critical point at $T = 0$ for $p\rightarrow p_{\rm{c}}$. 

\begin{figure}[ht]
	\includegraphics[width=0.4\textwidth]{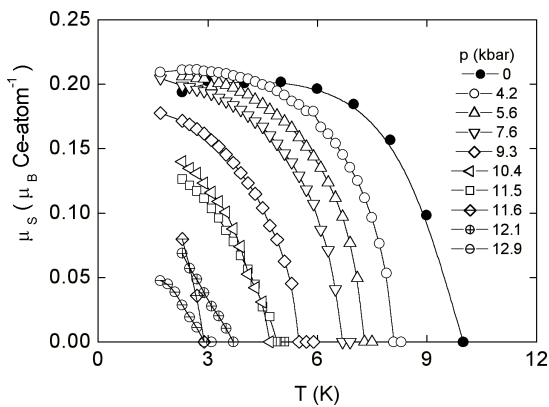}
	\caption{Temperature dependence of the spontaneous moment $\mu_{\rm{S}}$ for different pressures.}
	\label{figure7}
	\end{figure}

It is now interesting to establish if the suppression of the ordered magnetic moment as function of temperature is continuous at all pressures and whether the line $T_{\rm{C}}(p)$ (Fig.~\ref{figure6}c) is a line of second-order phase transitions.
Shown in Fig.~\ref{figure7} is the spontaneous moment $\mu_{\rm{S}}$ as function of temperature for various pressures in the range $0 < p < 12.9$\,kbar.
These data have been extrapolated directly from the hysteresis loops at various temperatures in the range $2.3$\,K\,$< T < 15$\,K, where an extrapolation by means of Arrott plots led to the same results. 
For all pressures at which a temperature dependence could be extracted from our data the magnetization vanishes continuously at $T_{\rm{C}}$.
The shallow maximum of $\mu_{\rm{S}}(T)$ seen at ambient pressure near $T^*\approx5$\,K disappears and is not present even at the lowest applied pressure of 4\,kbar.
When taken together this suggests that the line $T_{\rm{C}}(p)$ in the temperature-pressure plane corresponds indeed to a line of second-order phase transitions.

An irregularity that we cannot account for are variations in the initial slope of ${\rm{d}}\mu_{\rm{S}}/{\rm{d}}T$ just below $T_{\rm{C}}$, e.g., at 10.4\,kbar and 12.9\,kbar.
In particular, the pressure dependence of the ordered moment shown in Fig.~\ref{figure7} does not represent the strict zero-temperature limit. 
The variations in initial slope may in principle lead to considerable qualitative differences of the pressure dependence of $\mu_{\rm{S}}$ as compared to the temperature dependence shown in Fig.~\ref{figure2}a.
To resolve this issue requires further measurements to lower temperatures, which are outside the technical possibilities of our apparatus.

The pressure dependence of the Curie temperature $T_{\rm{C}}(p)$, shown in Fig.~\ref{figure6}c, has been determined from the data shown in Fig.~\ref{figure7} (filled squares) and slow temperature sweeps at a very low field of 0.01\,T (open circles; raw data is not shown). 
These temperature sweeps only served to check for anomalies that mark the Curie point, but cannot be analyzed properly further.
The solid line connecting the data points serves as guide to the eye. 
Down to 1.5\,K, the lowest temperatures accessible with our apparatus, $T_{\rm{C}}(p)$ is consistent with a quantum critical point at $p_c\approx13.1$\,kbar.

	\begin{figure}[hb]
	\includegraphics[width=0.4\textwidth]{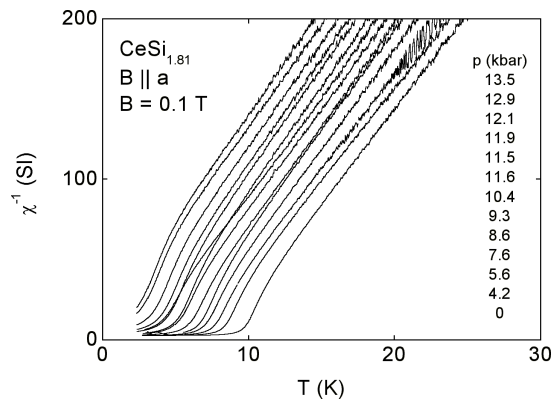}
	\caption{Inverse dc susceptibility as function of temperature for different pressures ($4.2$\,kbar $< p < 14.4$\,kbar) at $B = 0.5$\,T. The fast increase near $T_{\rm{C}}$ does not change with $p$. The curves are shifted to lower temperatures with increasing pressure.}
	\label{figure8}
	\end{figure}

Characteristic features of the inverse susceptibility as function of temperature at ambient pressure (Fig.~\ref{figure3}) are a Curie-Weiss dependence of large fluctuating moments at high temperatures and a strong downward curvature below about 20\,K.
These features characterize the environment out of which the ferromagnetic ground state at low temperatures emerges.
To interpret the suppression of the ordered moment as function of pressure it appears therefore important to track the general features of the susceptibility as function of pressure as well.

Shown in Fig.~\ref{figure8} are dc susceptibility data recorded in a magnetic field of 0.1\,T as function of temperature for various pressures.
Data have only been recorded up to 30\,K as the signal of the sample at higher $T$ is too small in comparison to that of the empty pressure cell to be resolved.
At high temperatures the slope of the susceptibility is essentially constant and the curves are shifted to the left (lower temperatures).
Further, even for temperatures approaching $T_{\rm{C}}$ the curves are remarkably parallel.
Deviations from being parallel for the lowest temperatures and highest pressures may be readily attributed to the polarizing effects of the applied magnetic field of 0.1\,T.

Our susceptibility data are consistent with a smooth evolution of the size of the ordered magnetic moments and unchanged second-order behavior with increasing pressure.
In particular the effective Curie-Weiss moment and the important energy scale marked by the downturn below 20\,K are essentially constant as function of pressure.
Pressure appears to affect only the initial susceptibility, defined as $\chi(T\to0,B\to0)$, which in present-day models of quantum phase transitions \cite{bel01,voj01} is used as the control parameter of the transition.
	
	\begin{figure}[ht]
	\includegraphics[width=0.4\textwidth]{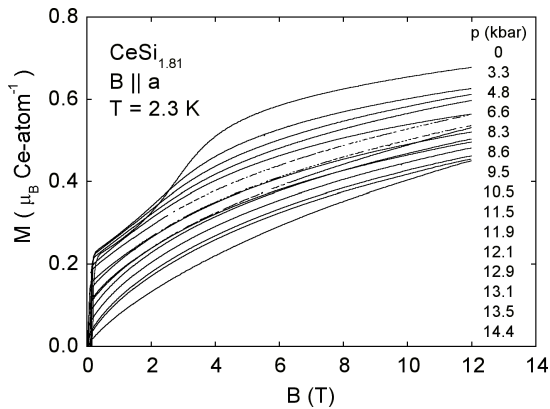}
	\caption{Magnetization under pressure as function of the magnetic field at $T = 2.3$\,K. As for $p = 0$ the field was directed along the $a$-axis of the single crystal. Applied pressures lie in the range between $0 < p < 14.4$\,kbar. The magnetization decreases monotonically with increasing pressure.}
	\label{figure9}
	\end{figure}

We finally turn to the pressure dependence of the high-field magnetization as measured at $T = 2.3$\,K for fields up to 12\,T.
Typical data are shown in Fig.~\ref{figure9}. 
With increasing pressure the curves are monotonically shifted to lower values and stay essentially parallel.
Thus the susceptibility at the highest fields, i.e. the slope of $M(B)$, is roughly constant as function of pressure, signaling that the magnetization is highly unsaturated.
The shoulder at $B^*\approx4$\,T which dominates $M(B)$ at $p=0$ vanishes almost completely under a pressure of only 4.2\,kbar and disappears altogether at higher pressures.
The suppression of the metamagnetic transition with increasing pressure is in line with the interpretation given above: 
The increased pressure strengthens the Kondo effect and thereby suppresses  the metamagnetic transition.
Despite a remarkable qualitative analogy with the variation of $M(B)$ at $p=0$ for increasing temperature, there is also an important difference.
Because the transition at $p=0$ as function of temperature is second order, $M(B)$ at 7.5\,K, i.e., just below $T_{\rm{C}}$, is highly nonlinear (Fig.~\ref{figure4}) and exhibits strong additional curvature over the entire magnetic field range up to 12\,T, in particular below $\sim3$\,T, in comparison to the $M(B)$ curves at other temperatures.
This non-linearity is expected for second-order behavior.
In contrast, the curvature of $M(B)$ as function of pressure is much less affected by the proximity to $p_c$. 

\section{Discussion}

The discontinuous suppression of ferromagnetism in pure ferromagnets \cite{pfle02,uhl04,nik05,pfle97} has led to the suggestion that for pure compounds ferromagnetism is \textit{always} suppressed discontinuously under pressure.
The conclusion drawn from the work on ZrZn$_2$ is related to the peculiar structure of the density of states near the Fermi level, notably there being a pronounced maximum in the vicinity of $E_{\rm{F}}$. \cite{uhl04,kim01}
The suppression of the magnetic order in weak transition-metal magnets as function of temperature at ambient pressure is essentially driven by a broadening of the density of states (DOS) due to thermal fluctuations.
Although hydrostatic pressure also acts to broaden the DOS via increased $d$-electron overlap, this transition in discontinuous, i.e., temperature and pressure suppress the magnetic order in a different fashion.

In contrast, the suppression of the easy-axis magnetization of {\cesi} under temperature and pressure shows a remarkable similarity.
In particular, the continuous suppression of the ordered magnetic moment appears to suggest that  {\cesi} is the first example of a genuine ferromagnetic quantum critical point.
However, four features in our magnetization data clearly establish that the magnetic state at ambient pressure and low temperatures is not that of a pure ferromagnet.
These features are: (i) a broad maximum in the ordered magnetic moment at $T^*\approx5$\,K, (ii), the emergence of an additional shoulder in the low-field magnetization below $\sim T^*$, (iii) the presence of metamagnetism at $B^*\approx4$\,T, and (iv) a broad maximum at a low pressure around $p^*\approx5$\,kbar.
In this respect, the role of Si deficiencies has to be considered.
Detailed four-circle $x$-ray diffraction studies of the crystallographic structure of samples from the same section of the ingot where the magnetization samples were cut has suggested an incommensurate superstructure of the Si vacancies. \cite{lei05}
As mentioned above, neutron scattering experiments do not confirm this superstructure and suggest that it may be a surface property. \cite{jano01}

Since all investigated ferromagnets exhibit a first-order transition when $T_{\rm{C}}$ approaches absolute zero, CeSi$_{1.81}$, appears at first sight as an exception. 
However, a possible first-order transition in this material might be smeared due to the Si lattice disorder.
The deviation from simple ferromagnetic behavior as described by the four features mentioned above may be due to an incommensurability of the Fermi surface introduced by the strain fields of an incommensurate structural modulation.
It may alternatively be due to variations of the magnetic moments of the Ce atoms from a local-moment point of view, that are introduced by incommensurate variations of the crystalline environment surrounding the Ce sites, keeping in mind that the Kondo effect arises essentially from the coupling of Ce 4f moments to the conduction-electron "configuration" in the local environment around a given Ce atom.
A possible metamagnetic transition induced by a magnetic field might also play a role.
An additional effect may be a change of the local crystalline electric field field.
In either case, Fermi-surface effect or local-environment picture, the density of states near the Fermi level is expected to display a complex structure that is clearly outside present-day models of pure ferromagnets.

By the same token, the effect of pressure in {\cesi} may be considerably more complex than just inducing a broadening of the conduction bands and associated suppression of the DOS near the Fermi level that leads to a reduction of $T_{\rm{C}}$, as in weak transition-metal ferromagnets.
Furthermore the proximity of the nonmagnetic $\alpha$-ThSi$_2$ phase has to be considered.
However, the nonmagnetic $\alpha$-ThSi$_2$ at large Si concentrations $x$ has a larger volume than the magnetically ordered $\alpha$-GdSi$_2$ phase
found for smaller $x$.
Therefore it is unlikely that pressure induces a structural change toward the nonmagnetic $\alpha$-ThSi$_2$ phase.
The incommensurate structural modulation also implies the presence of a modulation in the response to pressure and the build-up of inhomogeneous strains within the sample.
These internal concomitant strains may lead to smearing of the ferromagnetism and a suppression that is not accompanied by non-linearities in the magnetic-field dependence.

\vspace{5mm}

\section{Conclusions}

We have studied the easy-axis magnetization of {\cesi} as function of temperature and pressure.
The qualitative similarity of the temperature dependence at ambient pressure with the pressure dependence at low temperatures suggests that {\cesi} may be the first compound where a ferromagnetic quantum critical point may be induced by hydrostatic pressure.
However, the presence of an additional magnetic modulation at low temperatures and ambient pressure and the observation of a possible incommensurate superstructure of the Si vacancies questions a straightforward interpretation in terms of quantum criticality in pure ferromagnets.
Clearly, elastic neutron scattering, also under hydrostatic pressure, currently under way \cite {jano01}, is needed to disentangle the possible antiferromagnetic component which may be operative in this system, and its relation to the structure properties.

\section{Acknowledgments}
C. P. and S. D. acknowledge financial support in the framework of a Helmholtz-Hochschul-Nachwuchsgruppe of the Helmholtz Associations of Research Center under contract No. VI-NG-016.

\end{document}